\documentclass[twocolumn]{aastex63}
\newcommand{\T}{`T'}
\newcommand{\s}{`S'}
\newcommand{\Rsun}{$R_\odot$}
\newcommand{\mNCC}{$\langle NCC\rangle$}
\received{16-Dec-2020}
\submitjournal{\apjl}
\accepted{17-Feb-2021}
\shorttitle{Correlated modes in noise storm source evolution}
\shortauthors{Atul Mohan}
\graphicspath{{./}{figures/}}

\begin{document}
\title{Discovery of correlated evolution in solar noise storm source parameters: \\Insights on magnetic field dynamics during a microflare.}

\correspondingauthor{Atul Mohan}
\email{atulm@uio.no}

\author[0000-0002-1571-7931]{Atul Mohan}
\affiliation{Rosseland Centre for Solar Physics, University of Oslo, Postboks 1029 Blindern, N-0315 Oslo, Norway}
\affiliation{Institute of Theoretical Astrophysics, University of Oslo, Postboks 1029 Blindern, N-0315 Oslo, Norway}

\begin{abstract}
A solar type-I noise storm is produced by accelerated particle beams generated at active regions undergoing magnetic field restructuring. Their intensity varies by orders of magnitude within sub-second and sub-MHz scales. But, the morphological evolution of these sources are not studied at these scales, due to the lack of required imaging cadence and fidelity in metrewave bands. Using data from the Murchison Widefield Array (MWA), this work explores the coevolution of size, sky-orientation and intensity of a noise storm source associated with a weak microflare. The work presents the discovery of two correlated modes of evolution in the source parameters: a sausage like \s\ mode where the source intensity and size shows an anti-correlated evolution; and a torsional like \T\ mode where the source size and sky-orientation shows a correlated evolution. A flare mediated mode conversion is observed from \T\ to \s\ for the first time in these sources. These results support the idea of build up of magnetic stress energy in braided active region loops, which later go unstable causing flares and particle acceleration until they relax to a minimally braided state. The discovered mode conversion can be a future diagnostic to such active region phenomena.
\end{abstract}



\section{Introduction} \label{sec:intro}
{Solar type-I noise storms
are usually associated with active regions and sunspots,
during times of flaring or large scale magnetic field restructuring \citep{elgaroy1977, kathiravan07_stats_CMEdrivenTypeI,Iwai11_CME_enhancetypeI_link}. The bright radio emission is a result of coherent plasma emission mechanisms triggered by flare-accelerated electron beams trapped in active region magnetic field structures \citep{ginzburg1958,melrose1972}. However, there has been observations of noise storms, especially weak ones with flux enhancements typically less than 100 SFU (1 SFU = $10^{30}\,W\,m^{-2}Hz^{-1}$), which could not be associated with any particular flares \citep{smith62_flare_typeI_correlStats,lesqueren64_flare_typeINRH169locs}.
The deciding criteria for a noise storm to occur, and for a flare or active region to be linked to a noise storm are not well understood. Relatively recent works using multi-waveband data and sensitive modern radio arrays demonstrated that type-I sources can also be related to small scale magnetic enhancements and weak EUV brightening with no necessary flaring or major magnetic field restructuring \citep[e.g.][]{iwai12_typeI_smallscaleEUVmagDyn_link, Lin17_typeI_smallEUVflare_link, Suresh17_waveletbasedstudy, mohan2019b}.
Since the accelerated electrons beams driving the noise storm emission are produced at reconnection sites in these time varying magnetic field structures, their energy and spatial distribution functions are expected to evolve at similar scales \citep[e.g.][]{Gordov12_MagRelax_loops,Tomin10_e-acclEpisodes_typeI,Fyfe20_Forwmodel_MHDBraids}. These could leave observable signatures in the noise storm source morphology. 
However, to study the source dynamics in tandem with its sub-second and sub-MHz scale flux variability \citep[e.g.][]{wild1957_burstClassification,Elgaroy1970_typIcharacterisation,Guedel90-timeProfile_typIspikes,sundaram05_dt_df_typI}, high fidelity snapshot spectroscopic imaging at similar scales is essential.
This remained a challenge until the advent of modern interferometric arrays like Murchison Widefield Array \citep[MWA;][]{Tingay2013}, LOw Frequency ARray \citep[LOFAR;][]{vanHaarlem13} and the Long Wavelength Array \citep[LWA;][]{Ellingson13_LWA1}. 
A similar study was done on type-III bursts by \cite{mohan2019}. They reported the discovery of fast second-scale anti-correlated Quasi Periodic Pulsations (QPPs) in the sizes and flux density of type-III sources produced by a weak active region jet. The authors linked it to sausage modes in the active region supported by magnetic field modelling and Extreme Ultra Violiet (EUV) images of the jet. They also discovered pulsations in the source sky-orientation. 
}
This work will present the first study of simultaneous sub-second evolution of noise storm source parameters namely size, sky-orientation and integrated flux density. 
The event presented in \cite{mohan2019b} (hereafter M19) is chosen for this study since it is associated with a weak active region transient brightening \citep[ARTB;][]{shimizu92ARTB_yohkoh} with no major magnetic field re-structuring. Being weak, it can be assumed that several physical parameters remain practically unchanged due to the event, increasing the odds for discovering local MHD or plasma perturbative modes. 
Section \ref{sec:obs_analysis} describes the observations and image analysis. Section \ref{sec:discussion} discusses the emergent physical picture from the observed source evolution, followed by conclusions in Section \ref{sec:conclusions}. 
\section{Observations and analysis}
\label{sec:obs_analysis}
This study is based on archival data, recorded by the Murchison Widefield Array (MWA) Phase I on November 3, 2014 from 06:08:02 to 06:20:02 UT. The observation datasets had a bandwidth of 15.36 MHz, spectral resolution of 40 kHz and time resolution of 0.5s. Each observing session was 4 minutes long and centred at 199 MHz from 06:08:02 -- 06:12:02, 229 MHz from 06:12:02 -- 06:16:02 and again at 199 MHz 06:16:02 -- 06:20:02. 
The shift in the observation band was not intended for this study. 
An ARTB event occurred around the middle of the observation period, accompanied by a weak flare detected by RHESSI in 3--12 keV band. GOES satellites reported a simultaneous B6 class flare. 
The radio data hence covered the microflare from pre-flare to post-flare phase.
\begin{figure*}
    \centering
    \includegraphics[scale=0.2,width=0.8\textwidth,height=0.35\textheight]{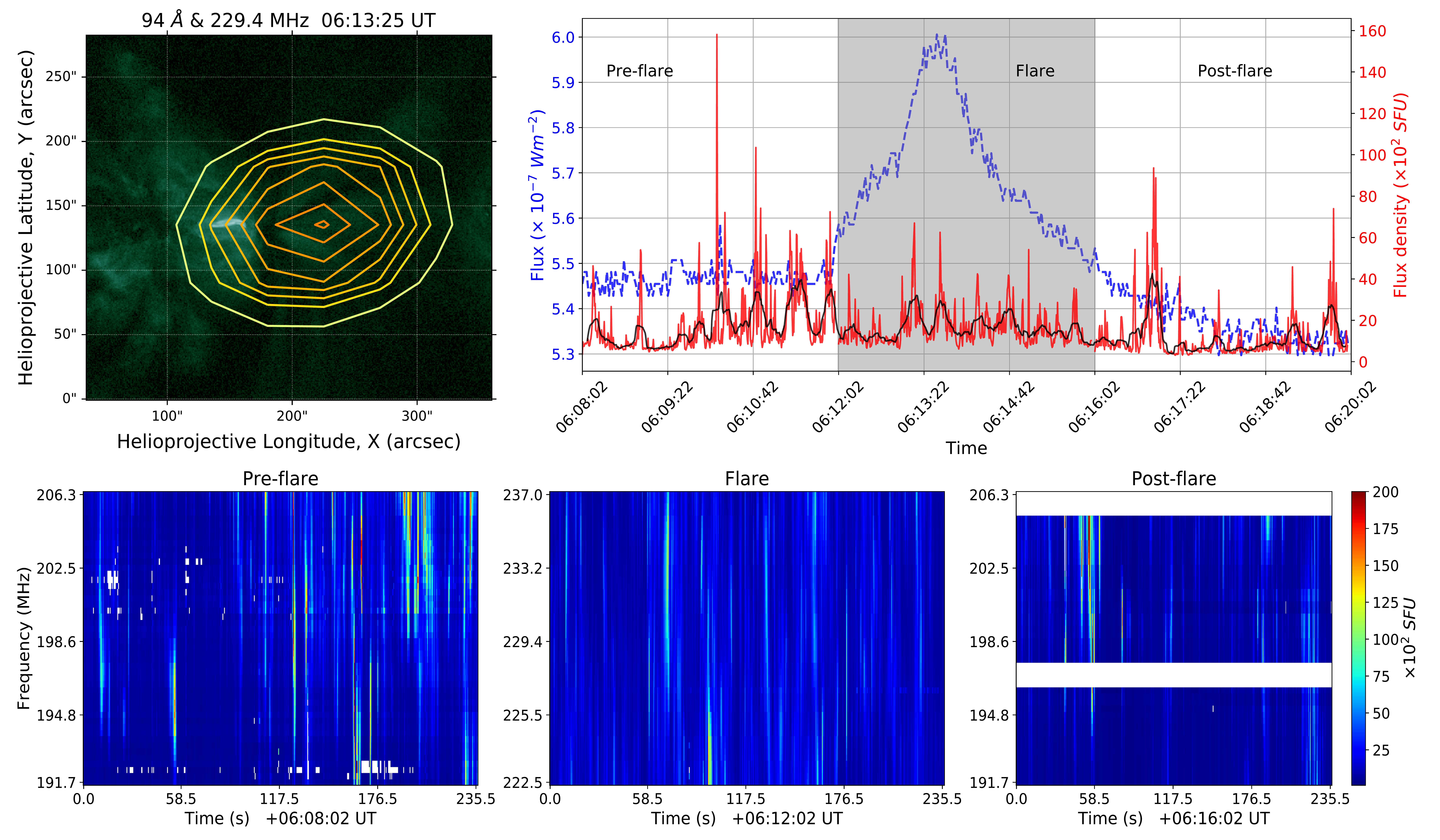}
    \caption{{\it Top row:} (Left) AIA 94\,\AA\ image of the Sun zoomed to the bright ARTB site. Overlaid are MWA 229\,MHz contours at 60, 70, 75, 80 ,85, 90, 95 \& 99 \% of the peak noise storm flux. Synthesized beam size: $3.9^{\prime}\times 2.9^{\prime}$. (Right) Red curve shows the spatially resolved band averaged light curve for the noise storm source. Black curve is a 20\,s running mean filtered light curve revealing the 30\,s QPPs. Overlaid in blue is the GOES 1 -- 8\,\AA\ light curve. The three event phases are demarcated. {\it Bottom row:} SPatially REsolved Dynamic Spectrum of the source during the three phases.}
    \label{fig:SPRETS_AIA_Img}
\end{figure*}
Imaging was done using the Automated Imaging Routine for Compact Arrays for the Radio Sun \citep[][]{mondal2019} at 0.5\,s cadence and 160\,kHz frequency resolution, using default parameters. The snapshot spectroscopic brightness temperature maps were made using these images following the prescription in \cite{oberoi2017} and \cite{Atul17}.
The noise storm source was resolved in all the images with a size greater the synthesised beam (beam) by $\approx$20\% on average and had a 2D Gaussian morphology.
The left panel of Fig.\ref{fig:SPRETS_AIA_Img} shows noise storm source contours overlaid on an AIA 94\,\AA\ image, during the flare. 
This source has an FWHM of $4.7^{\prime} \times 3.3^{\prime}$ along its principal axes, when the FWHM of the beam is $3.9^{\prime}\times 2.9^{\prime}$. The beam-deconvolved source (``true source" hereafter) has $\approx 37\%$ of the beam size.
Using the imfit task of Common Astronomy Software Applications (CASA; \citealp{casa}), a 2D Gaussian function plus a constant background was fit to the burst source region in the images across time and frequency. 
The fitted constant accounted for the quiet Sun background.
The beam was deconvolved from the best-fit Gaussian function to derive the true source dimensions: the major and minor axis widths ($\sigma_{major/minor}$); the position angle and the integrated flux density which is the total flux within the FWHM sized ellipse. 
Area of the true sources were estimated as $\pi \sigma_{major} \sigma_{minor}$. The errors on the best-fit parameters were appropriately propagated to calculate the errors in area. 
SPatially REsolved Dynamic Spectra (SPREDS) for the true source was made using the derived integrated flux density. This is shown in the bottom panels of Fig.\ref{fig:SPRETS_AIA_Img}. The intermittent white patches show regions where either the data were bad or the estimates were less than thrice their uncertainties. 
The top right panel of the figure shows a band averaged light curve for the source obtained from SPREDS in red. The black curve is obtained by applying a 20\,s wide running mean window. 
30\,s QPPs can be seen riding over a nonthermal flux floor, which rises during the flare. 
The GOES X-ray light curve in the 1--8\,\AA\ band is shown in blue. The data is divided into three phases based on the flare evolution: Pre-flare, flare and post-flare. 
Subsequent sections will focus on the spectro-temporal coevolution of the  morphological parameters of the noise storm source (area and position angle) in tandem with its integrated flux density during these phases.
Earlier studies usually approached the noise storm emission as a bright continuum, superposed with spiky burst features (type-I bursts) \citep[e.g.][]{Mercier97_typeI_nanoflare,iwai14_typeI_finefeature_hists,Suresh17_waveletbasedstudy}. 
This work will analyse the emission as a whole, from an active source varying its flux and morphology in tandem with the associated ARTB (microflare).
\subsection{Coevolution of source parameters}
\label{sec:parameter_evol}
\begin{figure}
    \centering
    \includegraphics[scale=0.3,width=0.45\textwidth,height=0.36\textheight]{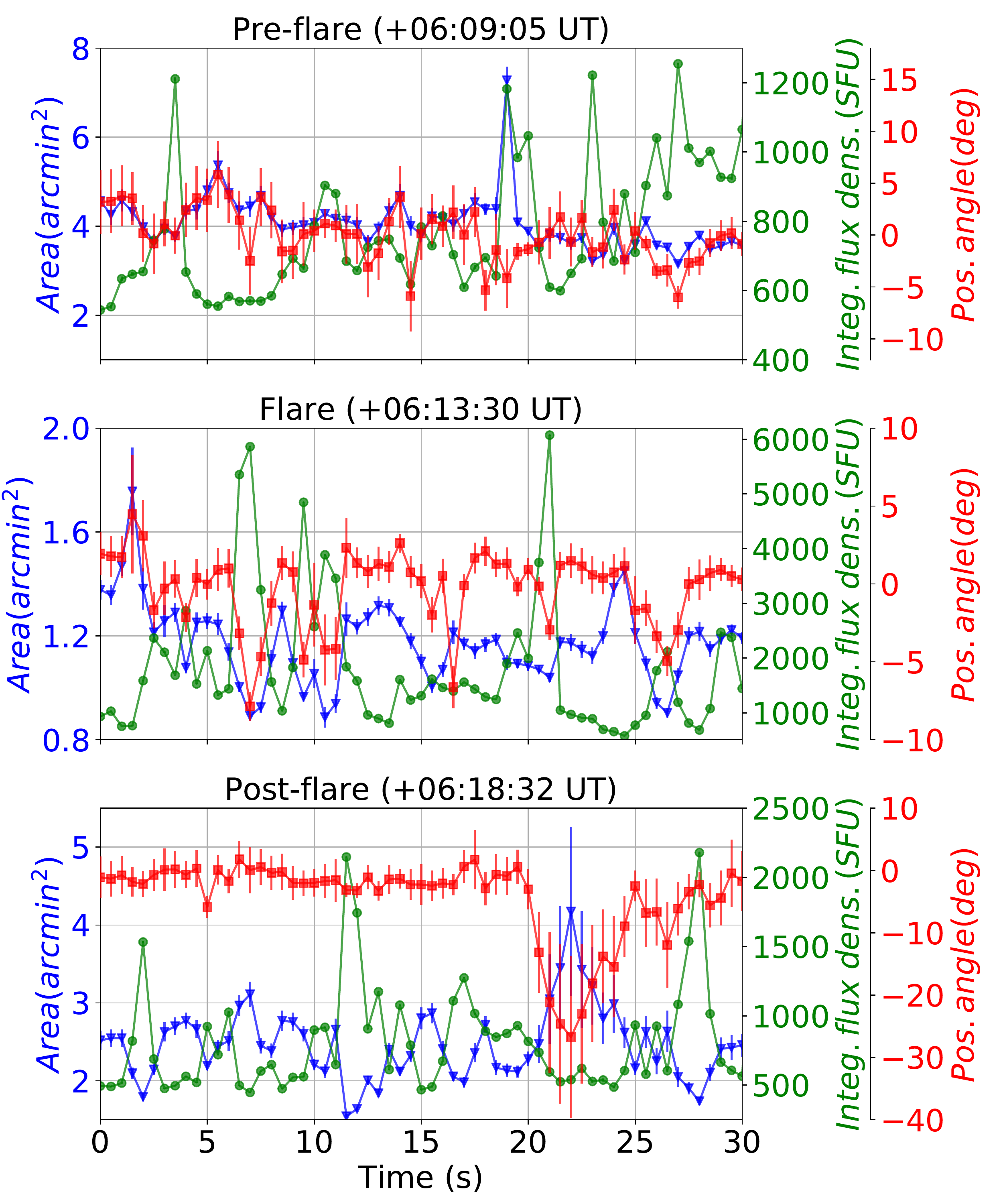}
    \caption{{Coevolution of integrated flux density, area and position angle during different phases, within a 30\,s window. The plotted data come from the mid frequency of the respective bands. Start time of each data is given in the title.}}
    \label{fig:coevol_area_flux_posang}
\end{figure}

Figure \ref{fig:coevol_area_flux_posang} presents the temporal evolution of integrated flux density and median-subtracted position angle (hereafter, position angle) of the source with its area during various phases.
The data presented for the pre-flare phase is from a period devoid of strong bursts; the flare phase sample data is from a period of intense bursts; and the post-flare phase data is from a period well after the radio flux dropped. 
The integrated flux density, area and position angle of the noise storm source show rapid variability with occasional strong pulses. These parameters also show periods of correlated evolution in all phases. 
For example, in the pre-flare phase, area and integrated flux density show an anti-correlation in the first $\approx 8\,s$. Later, they evolve correlated with a common peak around $18\,s$. Beyond $25\,s$ the floor of the integrated flux density rises steadily, but the area varies around a fixed floor. Their coevolutionary trend seems erratic in this phase.
\begin{figure}
    \centering
    \includegraphics[scale=0.3,width=0.48\textwidth,height=0.18\textheight]{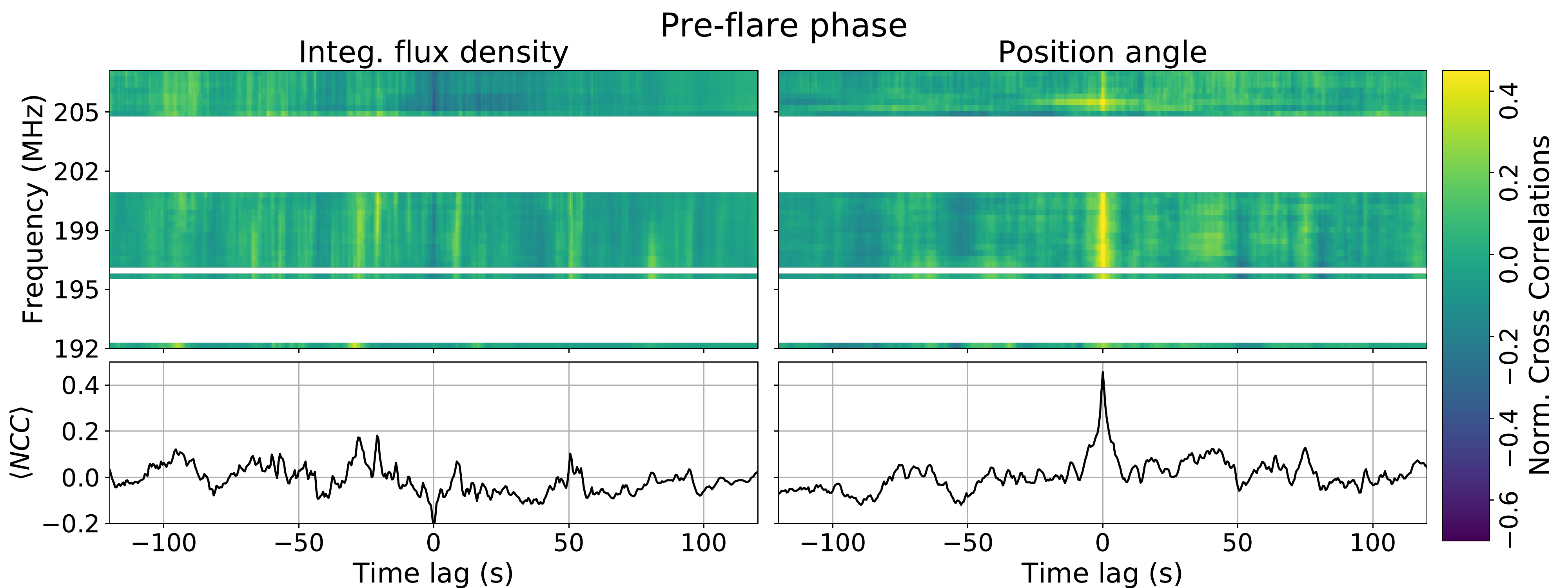}\\
    \includegraphics[scale=0.3,width=0.48\textwidth,height=0.18\textheight]{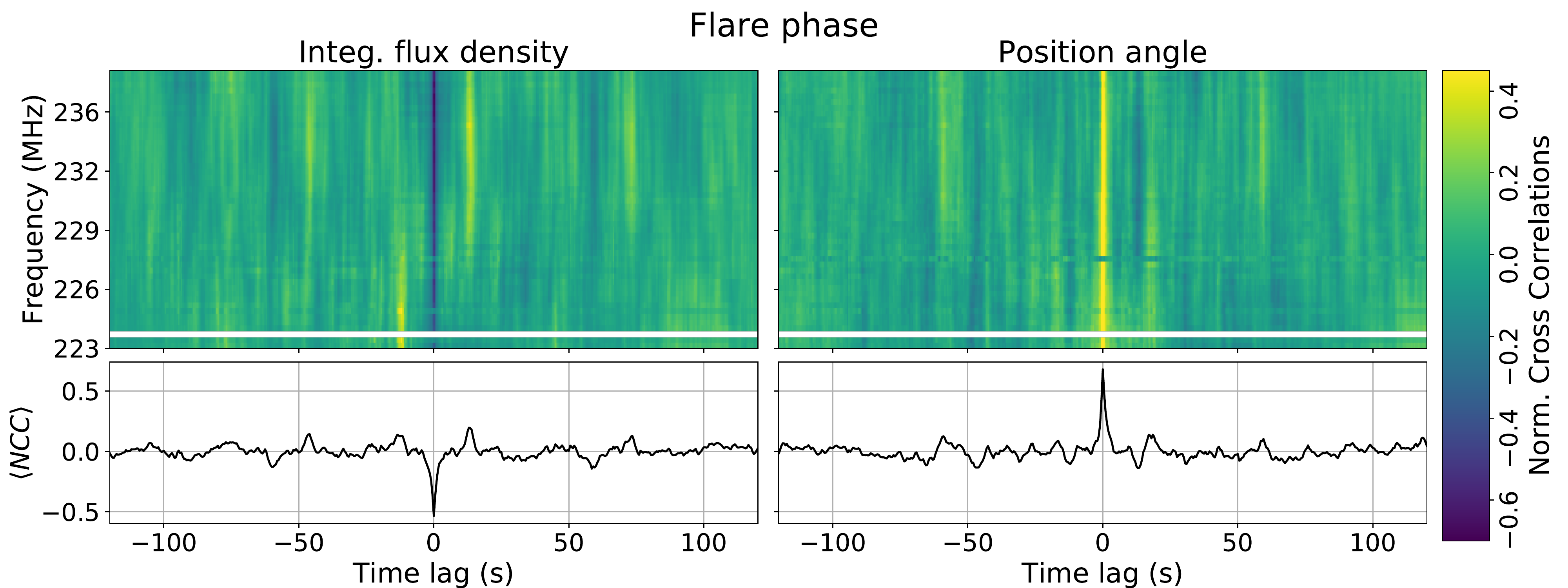}\\
    \includegraphics[scale=0.3,width=0.48\textwidth,height=0.18\textheight]{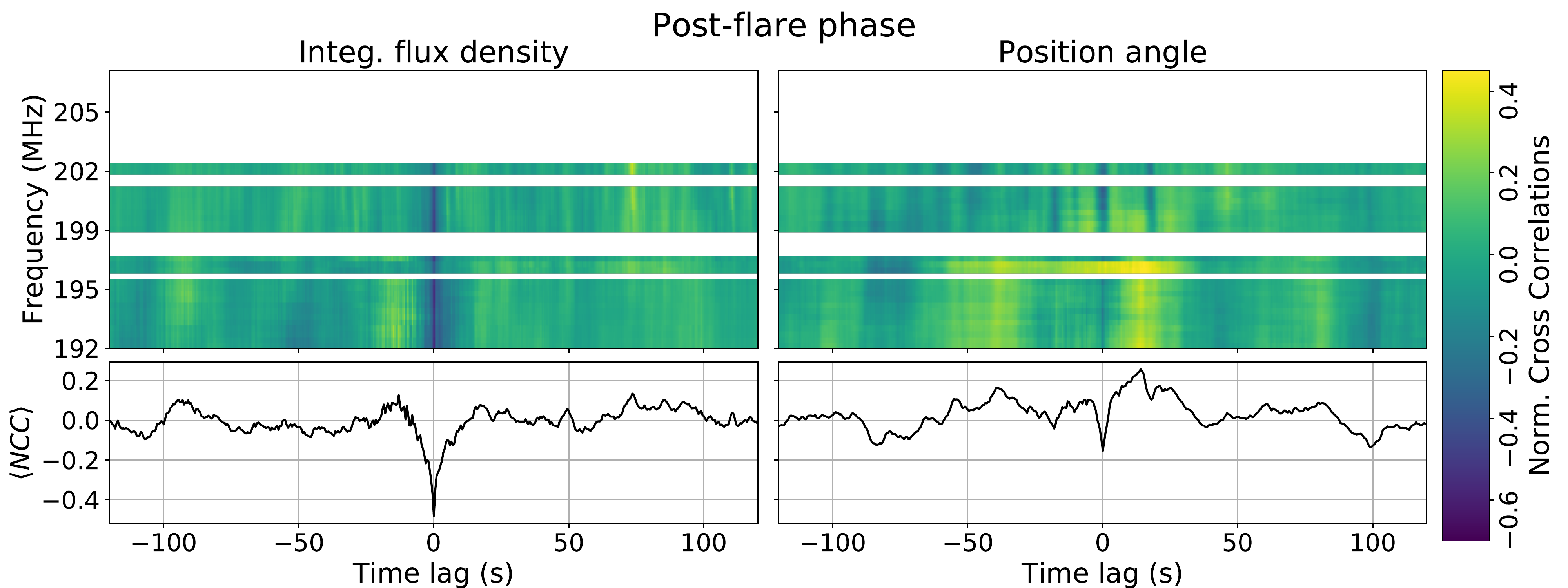}
    \caption{Matrix plots present the NCC functions for integrated flux density (left) and position angle (right) of the source with respect to its area, for every observation frequency during different phases. The masked bands had many data gaps. The line plot below each matrix plot is the band averaged NCC ($\langle NCC \rangle$).}
    \label{fig:NCC_area_flux_posang}
\end{figure}
Similar erratic nature is seen in position angle and area coevolution in the post-flare phase. When the first $\approx 10\,s$ give an impression that the two parameters are correlated, their trends  dissociate around $15\,s$ and become anti-correlated beyond $20\,s$.
However, in the flare and post-flare phases the anti-correlated evolution of area and integrated flux density is evident. Similarly area and position angle show a correlated evolution in the pre-flare and flare phases.
To get a clear picture of coevolution of these parameters in each phase, a normalised cross-correlation (NCC) analysis was carried out. Source area was chosen as the base parameter with which the others where correlated. 
NCC functions were evaluated for each pair of parameters, at every observation frequency, by correlating their full time profiles at 0.5\,s cadence in each phase.
Fig.\ref{fig:NCC_area_flux_posang} shows the NCC matrices truncated at $\pm$ 2\,minutes. The observation frequencies with significant number of masked data points were excluded in this analysis. The NCC functions are quite similar across the band. 
Hence a representative band averaged NCC (\mNCC) was computed. 
Assuming the coronal density model by \cite{zucca2014}, 15 MHz (30 MHz) band corresponds to a region less than 10\% (14\%) of the pressure scale height of the local corona. So, the mean physical and dynamical properties are expected to be similar across this band, as seen in the NCC functions.
Extending the same argument, since the central frequencies of the observation bands differ only by 30\,MHz, \mNCC\ for all phases belong to the same coronal region. 
This work presents the discovery of correlated evolution in the three ``independent" parameters that define a noise storm source. 
\section{Discussion}
\label{sec:discussion}
Analysis of the effects of radiowave scattering and imaging artefacts in the observed trends, confirm their noise storm source origin. It also revealed that, despite the coevolution, the values of the parameters show no definitive trends amongst each other (See Appendix \ref{ap:1}).

\begin{figure}
    \centering
    \includegraphics[width=0.48\textwidth,height=0.22\textheight]{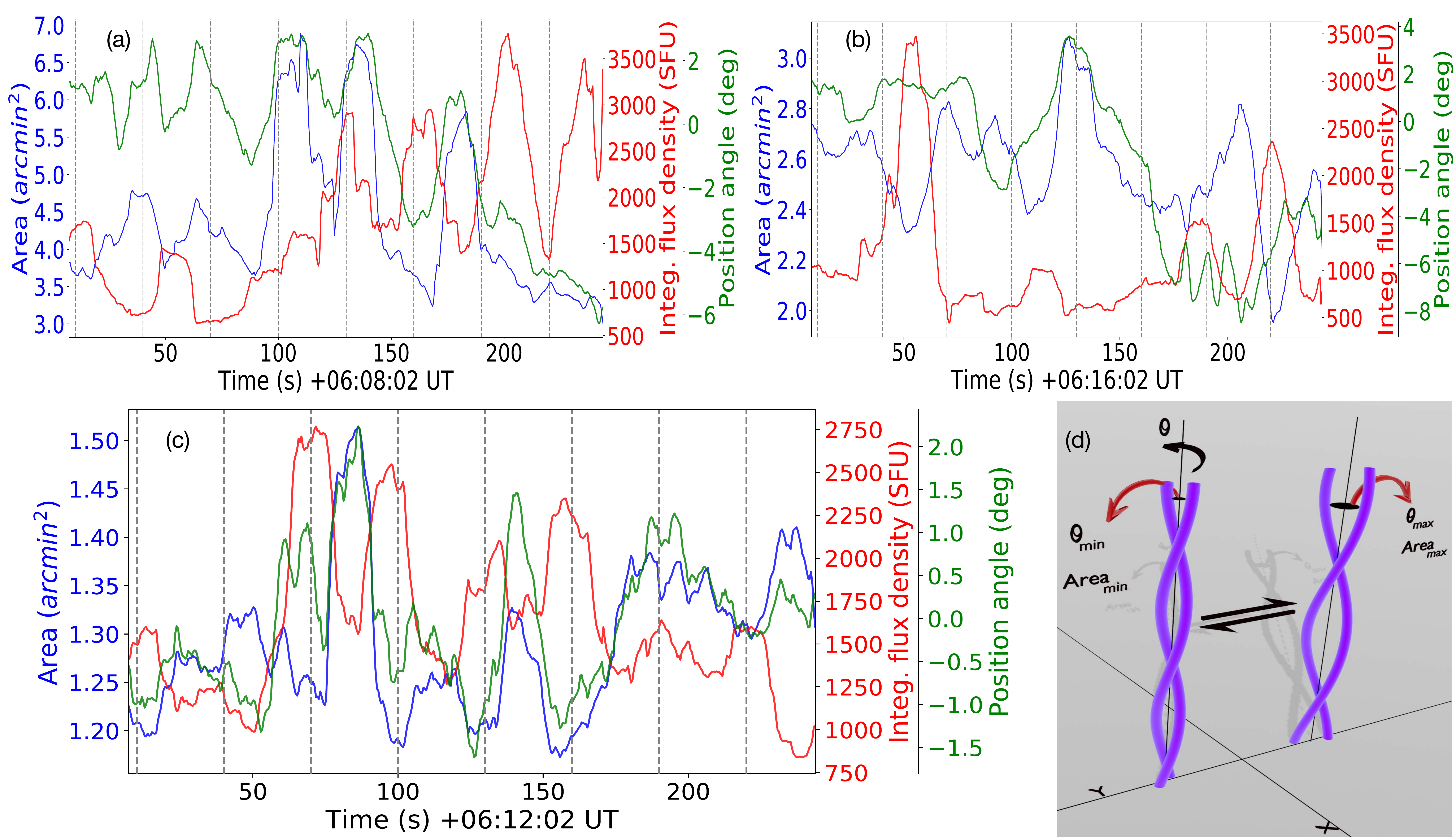}
    \caption{(a-c): Band averaged light curves of area, net flux density and position angle of the true source after applying a 30\,s running mean filter. Vertical lines are marked every 30\,s. (d): ``T" mode schematic showing the correlated evolution of area and position angle ($\theta$) as a braid wind/unwind. The subscript min/max indicates the relative parameter values.}
    \label{fig:30sQPPs}
\end{figure}
{M19 showed that the noise storm source is part of a large loop structure and is dynamically linked to a small active-region loop via a common magnetic footpoint (M19 Fig.11). The small loop underwent an ARTB during the flare phase, simultaneously enhancing the noise storm activity.}
The EUV analysis of the ARTB region revealed a braided structure at $\approx$ 12 Mm scale during the flare. 
The magnetic stresses continuously built up across the braid could have been released via enhanced particle acceleration events, causing the rise in radio flux.
The radio light curve showed 30\,s QPPs, which became more regular during the ARTB (Fig.\ref{fig:SPRETS_AIA_Img}).
The Alfv\'{e}n speed estimate (0.4\,Mm\,s$^{-1}$) from the magnetic field modelling and the QPP period provide a length scale of 12\,Mm in the radio source region. 
This matches the braiding scale at the ARTB site. So, M19 suggested that the radio source and the ARTB region are probably driven by their common footpoint, leading to the braiding of local loop structures at similar scales.
The bright type-I bursts seen in SPREDS are clumped within 30\,s periods which made the authors propose a periodic excitation of particle acceleration episodes like an avalanche within the local Alf\'{e}n timescale as the braided loops relaxed their excess internal energy continuously pumped in from below.
The absence of trends between any two source parameter values, support the picture of random particle acceleration episodes with differing scales as part of an avalanche. 
{Figure \ref{fig:30sQPPs} shows the evolution of area, integrated flux density and position angle of the true source, averaged across the band during different phases. The data were smoothed by a 30\,s running mean filter. Vertical lines are marked every 30\,s.
I report the discovery of 30\,s QPPs in the area and position angle of the noise storm source, in tandem with its integrated flux density.}
This strengthens the hypothesis of a braid that relaxes at Alfv\'{e}n timescales.
The correlated evolution of the true source parameters and the QPPs will now be analysed based on the above picture.
\subsection{Nature of correlated evolution \& its implications}
From Fig.\ref{fig:NCC_area_flux_posang}, it is inferred that there are two dominant modes of correlated evolution in source parameters:
an area-position angle correlated mode (\T\ mode hereafter) and an area-integrated flux density anti-correlated mode (\s\ mode hereafter).
The source area is a proxy to the size of the region of instability driven by accelerated electron beams produced at particle acceleration sites, that are magnetically linked to the noise storm source region. The source position angle is a proxy to the direction or tilt of the propagating beams and the integrated flux density relates to the beam energy flux density. 
So, \s\ mode can be envisaged as a sausage like mode where the area of the instability region and the energy flux density of the beam electrons are anti-correlated. Similarly, \T\ mode is akin to a winding-unwinding mode like the illustration in Fig.\ref{fig:30sQPPs}(d) where, size and orientation of the electron beam varies in a correlated manner as the braid switches from a tight to loose winding configuration. 
{Figure \ref{fig:30sQPPs}(a-c) shows 30\,s running mean filtered trends for each parameter. Simultaneous 30\,s QPPs are found in all parameters in every phase. 
These QPPs show a correlated evolution consistent with the dominant mode in that phase.
Interestingly, the type-I bursts which are seen clumped within the 30\,s periods also show the same coevolutionary behaviour (Fig.\ref{fig:coevol_area_flux_posang} \& Fig.\ref{fig:30sQPPs}).
This could be because the accelerated beams causing the bursts are jointly releasing the excess energy accumulated in some mode (\T\ and/or \s) during the Alfv\'{e}n (QPP) timescale, across the local dominant braid. 
The noise storm continuum could be comprised of numerous unresolved low energy bursts.
\T\ mode dominates in the pre-flare phase.
The flare phase marks the rise of \s\ mode alongside \T, which gives way for \s\ in the post-flare phase. This hints at a conversion in the dominant mode via the flare.}
The possible physical implications will now be discussed.
\subsubsection{Pre-flare phase}
The physical picture put forth by M19, suggest that the fast twisting motion in the magnetic strands driven by the footpoint motions causes the energy build-up primarily in the \T\ mode. Sausage like \s\ mode is absent in this phase.
The dominant braided structure goes unstable at Alfv\'{e}n timescale of 30\,s and release the excess energy via an avalanche of reconnection events producing accelerated electron beams. They cause the observed bursts with the \T\ mode imprinted. 
\subsubsection{Flare and post-flare phase}
The \T\ mode enhanced in the flare phase, possibly due to the twisting of already critically braided field structures. 
Simulations show that, this can cause kink instabilities in the loop and the excess energy gets released as accelerated particle beams and local heating, followed by gradual internal restructuring \citep[e.g.][]{Gordov12_MagRelax_loops, Threlfall18_interacting_twisted_loops}.
ARTB and the X-ray flare are signs of heating. The
radio flux hike could be due to increased particle acceleration events. The rise of \s\ mode during flare phase is noteworthy. All these hint at a redistribution of the excess \T\ mode energy to other degrees of freedom.
In post-flare phase the \T\ mode gives way to \s.
This mode conversion is possibly a sign of a restructuring loop.

Earlier studies on internally twisted loops targeted flares with strong hard X-ray and microwave emission, for measurements with good signal to noise ratio \citep[e.g.][]{Gordovs12_Effect_XrayLC_internalReconn,Sharykin18_Mclassflare_twistloopEmiss,Gordovs_20_XclassFlare_particleAcc}.
Here, a new way is presented to study such loops by tracking the dominant modes of evolution in the associated bright noise storm sources, and there by help bypass the constraint on flare energy.
\section{Conclusions}
\label{sec:conclusions}
The coevolution of source area, sky-orientation and integrated flux density of a noise storm, associated with an ARTB (microflare) is presented. This work presents the discovery of simultaneous and often correlated variations in these parameters. 
Correlated quasi periodic pulsations (QPPs) in area, position angle and flux density of the noise storm source are also discovered.
Normalised cross correlation (NCC) analysis between the parameters during the pre-flare, flare and post-flare phases revealed two dominant modes of correlated evolution: area-integrated flux density anti-correlated mode, named \s\ mode, like a sausage mode;  area-position angle correlated mode, named \T\ mode, like a winding-unwinding mode.
A conversion in the dominant mode from \T\ to \s\ is discovered, mediated by the flare. This can be a signature of the release of excess magnetic stress energy built up in \T\ mode in the local coronal loops, during the flare. Eventually, the \T\ mode energy density is redistributed to \s\ mode and particle energy. Such mode evolution patterns in associated noise storm sources can be used as diagnostics to study the evolution of flaring loops, regardless of flare energy.\\
\textbf{Acknowledgements: }This scientific work makes use of the Murchison Radio-astronomy Observatory (MRO), operated by the Commonwealth Scientific and Industrial Research Organisation (CSIRO).
We acknowledge the Wajarri Yamatji people as the traditional owners of the Observatory site. 
Support for the operation of the MWA is provided by the Australian Government's National Collaborative Research Infrastructure Strategy (NCRIS), under a contract to Curtin University administered by Astronomy Australia Limited. We acknowledge the Pawsey Supercomputing Centre, which is supported by the Western Australian and Australian Governments. 
This work is supported by the Research Council of Norway through its Centres of Excellence scheme, project number 262622 (``Rosseland Centre for Solar Physics'').  
AM acknowledges support from the EMISSA project funded by the Research Council of Norway (project number 286853). AM acknowledges Olga Mohan for the graphical support. AM acknowledges Prof. Divya Oberoi, Surajit Mondal and the anonymous referee for useful discussions. 
This research made use of NASA's Astrophysics Data System (ADS). 
\facilities{MWA, SDO(AIA), RHESSI and GOES}
\software{Numpy \citep{numpy},
        Astropy \citep{2013A&A...558A..33A},
          Matplotlib \citep{matplotlib},
          CASA \citep{casa},
          Sunpy \citep{sunpy}}

\appendix
\vspace{-1cm}
\section{Analysis of the effects of scattering \& imaging artefacts}\label{ap:1}
The effect of radiowave scattering and the possible imaging artefacts in the observed parameter  evolution will be discussed here. 
Scattering changes the absolute source size, as a convolution by a Gaussian scatter function, the width of which depends on the mean statistical properties of the ambient plasma \citep{Arzner1999,kontar2017_NatCom,mohan2019}. The noise storm source region is located at around 1.14 \Rsun\ when the ARTB source was at $\approx$ 1.02 \Rsun (See Fig 11. in M19). Though magnetically connected, the regions are spatially so far apart for the ARTB to have varied the ambient density fluctuation index ($\delta N/N$) at the noise storm region, sufficient enough to cause the observed large fractional changes in its area by about a few to $\approx 100\%$ within spans of a few second (Fig\ref{fig:coevol_area_flux_posang}).
Hence, the role of scattering in the observed source variability can be discarded.
The noise storm sources are well described by a single 2D Gaussian morphology. The low errors in the derived quantities, shown in Fig.~\ref{fig:coevol_area_flux_posang}, testify this. The errors in position angle appear large, as the plotted data are median subtracted.
However, in reality there could be multiple sources which are unresolved by MWA and(or) are smeared by scattering. But, this study focuses on the overall effective size and shape of the radio source which is well captured by the Gaussian fitting procedure.

Though MWA is a dense compact array with 8128 baselines within 100 - 3000 m range \citep{Tingay2013}, since the true source sizes are smaller than the beam extent, any possible systematic effects due to beam structure variations or angular resolution should be studied. Imaging at every time and frequency bin is an independent process with a unique beam shape. Still, the true source parameter trends look similar across frequency for the entire observation period. This increases the odds of them being intrinsic.
\begin{figure*}
    \centering
    \includegraphics[scale=0.3]{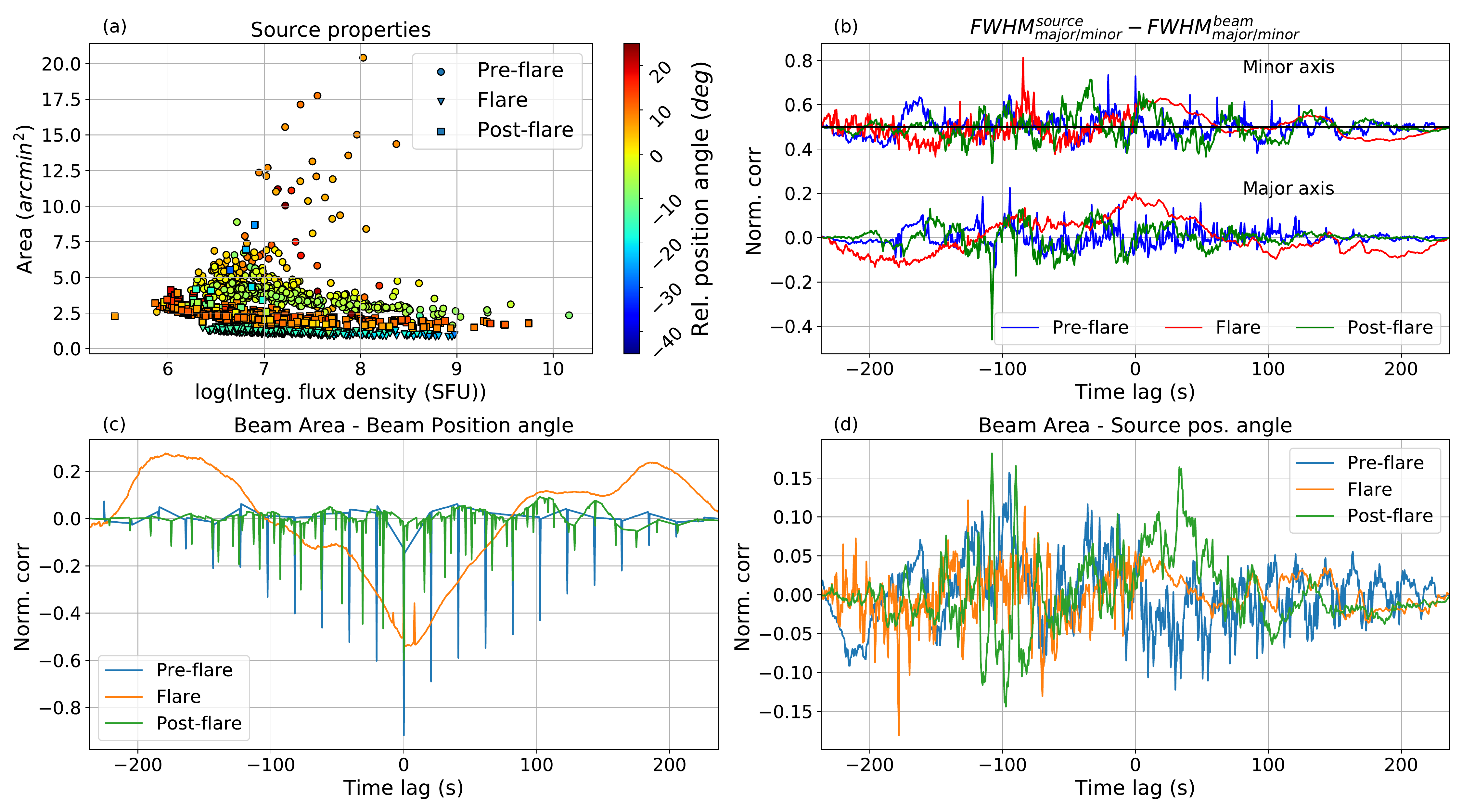}
    \caption{(a): Area versus integrated flux density of the true source. The position angle of the true source relative to the beam is color coded. (b):NCC function for the true source and the beam minor axes widths (Above) and major axes widths (Below). NCC values for minor axes are shifted by 0.5 for visual clarity. (c-d): NCC functions for the area and position angle of the beam (c), and for the beam area and the true source position angle (d).}
    \label{fig:beam_effects}
\end{figure*}
Figure \ref{fig:beam_effects}(a) shows various analysis done for the mid-band data during various observation phases. The choice of mid-band is arbitrary and it is expected that the imaging artefacts, if any, due to the issue of source size should be of the same statistical nature at all frequencies. This is due to the similarity in the source sizes and imaging procedure across the band. Color coded are the true source position angle at each instant after subtracting the corresponding beam position angle.
There is no significant trend among parameters, especially in the mid ranges of their values which are unaffected by any bias due to low event counts. 
If some systematic effects had seeped into the true source structure due to its relatively small size, the true source dimensions derived would have systematically mimicked the beam with relative position angle tending to zero towards small area values.
Figure \ref{fig:beam_effects}(b) shows the NCC for the FWHM of minor axes of the beam and the true source in the top. NCC values are hiked by 0.5 for representative purposes and the black horizontal line at 0.5 marks the true NCC=0 line. Similar analysis for major axes is shown below. There is no correlation between the angular scales of the true source and the beam even in the flare phase, when source sizes were relatively smaller. 
Figure \ref{fig:beam_effects}(c) shows the NCC for the area and position angle of the beam. There is no sign of correlation during any observation phase, unlike the true source data. On the contrary, it shows spikes in pre-flare and post-flare curves and an anti-correlation prominent in flare phase.
These spikes resulted since the imaging pipeline performed fresh calibration runs on the data every $\approx$ 20\,s leading to different antenna flagging schemes, which affected the beam structure. The solution from each calibration run was applied to make images in the intermediate time steps. In the flare phase, since source is very bright, high dynamic range images could be obtained with just a few rounds of self calibration after applying calibration solutions from initial time slice.
Figure \ref{fig:beam_effects}(d) shows the NCC between beam area and true source position angle which shows no sign of correlated evolution in any phase, unlike the NCC of the true source data.
These results increase the confidence in the observed true source area - position angle trends and assures that they are not beam-driven. 
\bibliography{main.bib}{}
\bibliographystyle{aasjournal}


\end{document}